
%
%
\magnification=1200
\pretolerance=10000
\hsize=4.5 in
\hoffset=1.0 cm
 \overfullrule = 0pt
\line{ }
\vskip 0.5 truecm
\rightline {LNF 92/110 P December 1992}
\vskip 2. truecm
\centerline{\bf LARGE N LATTICE QED.}
\vskip 2 truecm
\centerline { V.~Azcoiti }
\vskip 0.15 truecm
\centerline {\it Departamento de F\'\i sica Te\'orica, Facultad de
Ciencias, Universidad de Zaragoza,}
\centerline {\it 50009 Zaragoza (Spain)}
\vskip 0.5 truecm
\centerline { G. Di Carlo and A.F. Grillo }
\vskip 0.15 truecm
\centerline {\it Istituto Nazionale di Fisica Nucleare, Laboratori
Nazionali di Frascati,}
\centerline {\it P.O.B. 13 - Frascati (Italy). }
\vskip 3 truecm
\centerline {ABSTRACT}
\vskip 0.3 truecm

We study the $\beta, N$ critical behaviour of non compact QED with $N$ species
of light fermions, using a method we have proposed for unquenched simulations.
We find that there exist two phase transition
lines: one, second order, and the other, first order, that approaches
asymptotically the $\beta=0$ axis.
These two lines have different physical origin, the second one
being entirely due to fermions effects. We discuss the effect of the
approximation used, in terms of an expansion of the effective action in powers
of $N$, and conclude that the general features should not be
affected by this approximation.
\vskip 3 truecm

\vfill\eject

\par
\noindent

Strongly coupled lattice QED has been, in the last years,
 the object of much attention,
both analytical and numerical. The main subject of
these investigations has been the question
 if a strongly coupled abelian model
can be constructed with non trivial interactions in the continuum limit.

During some time, the results about the nature of the fixed point
of this model were controversial in the sense that
different groups reported different values of critical exponents
respectively consistent or inconsistent with a mean field description
of this model. More recently, an extensive investigation of the
quenched model in large lattices and at small fermion masses [1,2]
gave
evidence that a precise determination of the critical coupling $\beta_c$
is needed in order to extract critical indices with good accuracy,
since
small variations in the value of $\beta_c$ may induce strong changes in the
values of the critical exponents towards their mean field values [3].

The results of [1,2] reporting
non mean field values for the critical exponents in
the quenched model, have been improved for the unquenched case in [4,5]
and also by us in [3,6] in a completely independent calculation which
allowed the numerical determination of $\beta_c$ from the results of the
plaquette energy without the need of any kind of
extrapolation at zero fermion mass.
Additionally, an analysis, based on Renormalization Group  approach, performed
in [6] on the
results for the fermionic effective action,
pointed again to
a non gaussian nature for the strongly coupled fixed point of this model.
These results are in contrast with those in [7].

The agreement between the results of refs. [4,5] and those of refs. [3,6]
for the two and four flavour cases was certainly encouraging and
suggested the existence of a non trivial continuum limit for strongly coupled
QED, at least if the number of dynamical flavours is less or equal than four.

Another intriguing result reported in [5] was that critical indices for the two
and four flavours models, obtained from numerical simulations and the use of
the equation of state, were
found compatible between them and also with the critical
exponents of the monopole percolation transition, thus suggesting that
different flavour models are in the same universality class.

This result can be also understood in our approach, if the critical
value of the plaquette energy is independent on the number of flavours, as
suggested by the results of [3]. In such a case, the value (for instance)
of the
$\delta$ exponent which measures the response of the system to an external
symmetry breaking
field will be given by the dominant contribution to the expansion
of the chiral condensate in powers of the number of flavours $N$ at fixed
pure gauge energy $E_c$ [3], and it should be independent of $N$.

The aim of this letter is to analyse in detail the dependence on
the number of flavours $N$ of the physical results and in particular to
investigate the phase diagramm of this model in the $N$, $\beta$ plane with
special attention to the analysis of the physical origin of the different phase
transitions observed.

The approach we use to simulate noncompact QED with dynamical fermions is that
of refs.[3,6], originally tested in the compact model [8], and based on the
introduction of an effective fermionic action $S^F_{eff}(E,N,m)$
which depends on the
pure gauge energy $E$, fermion mass $m$ and number of flavours $N$, and which
is related to the gauge fields $A_\mu(x)$ by the relation [3]

$$  e^{-S_{eff}^F(E,m)} \equiv <\big(\det \Delta\big)^{N\over 4}>_E = $$
$$ {\int [dA_{\mu}(x)] (\det
\Delta(m, A_{\mu}(x)))^{N\over 4}
\delta({1\over 2} \sum_{x,\mu < \nu} F_{\mu \nu}^2(x) - 6VE)
\over
\int [dA_{\mu}(x)]
\delta({1\over 2} \sum_{x,\mu < \nu} F_{\mu \nu}^2(x) - 6VE)}
\eqno(1)$$

\noindent
where $\Delta(m, A_\mu(x))$
is the fermionic matrix (we use staggered fermions), $E$ is
the normalized pure gauge energy and the denominator in (1) is the density of
states which can be analitically computed

$$ N(E)= C_G E^{{3\over2}(V - 1)} \eqno(2)$$

\noindent
$C_G$ in (2) is some irrelevant divergent constant and $V$ is the lattice
volume.

After the definition of the effective fermionic action, the partition function
of this model can be written as a one-dimensional integral

$${\cal Z} = \int dE N(E) e^{-6 \beta V E-S_{eff}^F(E,N,m)} \eqno(3)$$

\noindent
from which we can define an effective full action per unit volume as

$$ \bar S_{eff}(E,\beta,N,m) =
-{3\over 2} \ln E + 6 \beta E + \bar S_{eff}^F(E,N,m)  \eqno(4) $$

\noindent
$\bar S^F_{eff}(E,N,m)$
in (4) is the effective fermionic action (1) normalized to the
lattice volume.

The thermodynamics of this system can be studied now by means of the saddle
point technique. The mean plaquette energy $<E_p>=E_0(m,\beta,N)$
will be given by the solution of the saddle point equation [3]

$${1 \over {4E}}-\beta -{1\over 6}{\partial\over
\partial E} \bar S^F_{eff}(E,N,m)= 0 \eqno (5) $$

\noindent
satisfying the minimum condition

$${1 \over 4E^2}+{1\over 6}{\partial^2\over
\partial E^2} \bar S^F_{eff}(E,N,m) > 0 \eqno (6) $$

By differentiating
equation (5) respect to $\beta$ we get for the specific heat

$$ C_{\beta} =  {\partial \over \partial \beta} <E_p>=
 - \{{1\over 4E^2_0(m,\beta,N)}+{1\over6}{\partial^2\over\partial E^2}
\bar S^F_{eff}(E,N,m)\Big|_{E_0(m,\beta),N} \}^{-1}\eqno(7)$$

The effective fermionic action $\bar S^F_{eff}$
was computed in [3,6] as a power expansion on the flavour number $N$.

The observation of two different regimes in the behaviour of $\bar S^F_{eff}$
as a function of $E$, linear
in the small energy region and nonlinear at large energies,
and a carefull analysis of the numerical results allowed us the
determination of the critical value of the energy (which turned out to be
independent on the
number of flavours $N$) and of the critical coupling
$\beta_c$.
Furthermore it was established in [3] that the results for $\bar S^F_{eff}$
reported
in Fig.1 can be very well fitted by two polynomials with a gap in the second
energy derivative of $\bar S^F_{eff}$
at $E=E_c$ which manifests itself in the
specific heat $C_\beta$ as a second order phase transition.

The origin of this  non analyticity in the effective fermionic action
is not clear  at present.
However, our numerical results suggest that this
singular behaviour could have the same origin as the
monopole percolation transition of the quenched
model extensively analysed in [9]. In fact if we take our value of the
critical energy $E_c=1.016(10)$ (independent on the
number of flavours) and consider the zero flavour limit, we get
$\beta_c=0.246(2)$ in very good agreement with the
critical $\beta_c$ obtained
in [9] from the results for the monopole susceptibility in the quenched model.

Going back again to expression (7) it should be noticed that a non analyticity
of the effective fermionic action is not the only way to get a discontinuity
in the specific heat.
A discontinuity in $C_\beta$ can in fact be produced by a
zero in the
denominator of (7) through exact cancellation of the two terms in
this expression [3]. To this end, a negative value of the second energy
derivative of the effective fermionic action is necessary. But this is just
what happens in the large energy region as it can be deduced from the results
for the effective fermionic action reported in Fig.1.

The above arguments can be made quantitative and at the same time the
critical number of flavours $N_c$ can be simply estimated.
We
start from the cumulant expansion of the effective fermionic action [3]

$$-S^F_{eff}(E,N,m)= {N\over 4} <\ln \det \Delta(m,A_{\mu}(x))>_E $$
$$+ {N^2\over 32} \{<(\ln \det \Delta)^2>_E-<\ln \det \Delta>_E^2 \} + ...
\eqno(8)$$

\noindent
where $<O>_E$ means the mean value of the operator $O(A_\mu(x))$ computed
with the probability distribution $[dA_{\mu}(x)]
\delta({1\over 2} \sum_{x,\mu < \nu} F_{\mu \nu}^2(x) - 6VE)$.
The numerical results for the
succesive terms in the expansion (8) in a $8^4$ lattice and massless fermions
have been reported in [3], and are here extended to larger energies.
As a first approximation, we will consider only
the first contribution to (8), which is reported in Fig. 1 for $N=4$.

Fitting the results for the mean logarithm of the fermionic determinant
at $m = 0$ by
two polynomials: first order for $E<E_c=1.016$ and fifth
for $E>E_c$ we get
for the second energy derivative of the effective
fermionic action

$$ {\partial^2\over\partial E^2}\bar S^F_{eff}=0
\qquad\qquad\qquad\qquad\qquad\hfill (E\le E_c)\eqno(10)$$

$${\partial^2\over\partial E^2}\bar S^F_{eff} =
{N\over4} [-1.669 + 3.877 E - 2.494 E^2 +
0.494 E^3] \qquad  (E_c<E<2.5)$$

A simple analysis of these results tell us that in order to compensate the
pure gauge contribution to the denominator of the specific heat in (7) we
need $N=13.1$. If $N$ is large but less than $13.1$, the height of the peak
increases with $N$ and just at this critical value, the specific heat diverges.
Now, if we increase again the value of $N$, there will be an energy interval
where the denominator of the specific heat in (7) will be negative and
therefore no solutions of the saddle point equations (5), (6) will exist in
this energy interval. This means that these energies will not be accessible
to the system and hence, for $N>N_c$ a first order transition
will appear.

The phase diagram of massless noncompact QED in the $N$, $\beta$
plane which emerges from our results is plotted in Fig.2. The
continuous (broken) lines represent first (second) order phase transitions
respectively. The end point of the first order phase transition line is a
second order phase transition point with a divergent specific heat.
The first order line ends at some finite $\beta$ since
${\partial^2\over\partial E^2}\bar S^F_{eff}=0$ for $E<E_c$.
On the other hand, the second order line merges, for large $N$,
 into the first order one since
$E_c$ falls into the energy interval not accessible to the system, which
widens as $N$ increases.

Of course the quantitative results of this simple analysis could change if
we take into account higher order contributions to the effective fermionic
action (8), but the physics behind these phase transitions can be understood
at this simple level. What is independent on approximations is the fact that
there is no first order phase transition at $\beta=0$ for any finite
value of $N$.

This result can be rigorously proved since
the effective fermionic action $\bar S^F_{eff}(E,m,N)$ is bounded for
any finite $N$. Therefore, when $\beta$ goes to zero,
the effective action (4) has only one minimum at
$E=\infty$, whereas for any finite value of
$E$ the effective action is finite. Thus, at $\beta=0$ there is no first
order transition for any finite $N$.

A stronger result can be proved, under the very natural
assumption (corroborated by the experimental data) that the
effective fermionic action is monotonically increasing with $E$, namely that
at $N \to \infty$ $<E> = 0$ for all $\beta$. This implies that, in this limit
only one phase (Coulomb) does exist.

In fact, since the value of the fermionic determinant is bounded, it follows
that

$$ e^{N<\log\det\Delta>}\le e^{-S_{eff}^F(E,m)}
\le e^{N\log \Delta_{max}} \eqno(11)$$

\noindent
which implies that the
fermionic effective action diverges linearly with $N$.

As a consequence, for $N \to \infty$
$$ \bar S_{eff}\Big|_{E={1\over N}} \approx
{3 \over 2} \ln N + N K(E,m)\Big|_{E=0} \eqno(12) $$

\noindent
and

$$ \bar S_{eff}\Big|_{E=E_0} \approx
N K(E_0,m) \eqno(13) $$

The monotonicity of $\bar S_{eff}^F(E)$ then implies that
$$ \lim_{N \to \infty}\Big( \bar S_{eff} \Big|_{E=E_0} -
\bar S_{eff} \Big|_{E={1\over N}} \Big) = \infty \qquad \forall E_0 \ne 0
\eqno(14)$$

\noindent
implying that $E=0$ is the absolute minimum of $\bar S_{eff}$ for any
$\beta \ne 0$.
This, together with the previous discussion, implies that the first order
line reaches asymptotically the $\beta=0$ axis.

To compare with published results on this subject,
Kondo, Kikukava and Mino [10] found, within
the Schwinger-Dyson approach, a continuous phase transition line which
approches the $\beta=0$ axis asymptotically. The general discussion of the
above paragraph excludes such a behaviour for large $N$.

On the other hand Dagotto, Kocic and Kogut [11], in the framework of a
numerical simulation, found evidence for a second order transition line wich
becomes first order at large $N$,
crossing the $\beta=0$ axis
($N_c\sim 30$ at $\beta=0$). The probable origin of the disagreement between
this result and ours is that the
critical $\beta$ at $N=30$ is so small that it is difficult to distinguish it
from zero in a numerical simulation where metastability signal can be
observed also at $\beta=0$ [11].

{}From a physical point of view, these two phase transition lines have a
different origin. The second order line, as pointed before,
could have the same
origin as the monopole percolation transition of the pure
gauge model [5,9] . Indeed our results for the critical energy and
their independence
on the flavour number $N$ favour this interpretation.
The first order line is on
the other hand produced by pure fermionic effects. When the number of dynamical
fermions increases, the fermionic contribution to the denominator of the
specific
heat becomes more and more important and has the correct sign to cancel the
pure gauge contribution.

To finish let us say that these results have been also confirmed by our
numerical simulations of this model. In these numerical simulations we
have taken into account the first  contribution to the effective
fermionic action (8). In our opinion, the contribution of higher powers in
$N$ is not likely to change the qualitative behaviour depicted above.

In fact, since the effective fermionic action diverges linearly with $N$,
is bounded as $E\to\infty$ and
is  monotonically increasing, then,  barring
very peculiar behaviours, it is convex for large $E$ (corresponding
to $\beta \to 0$). In such a case, for $N$ large
enough the two terms in (7) can always compensate, so a first order
transition line will be in general present, at
large $N$ and small $\beta$.
In conclusion, although the numerical structure of the phase
diagram might vary, it is very likely that the qualitative features remain
unchanged.

This work has been partly supported through a CICYT (Spain) -
INFN (Italy)
collaboration.

\vfill
\eject

\line{}
\centerline {\bf REFERENCES}
\vskip 1 truecm
\item {1.} {E. Dagotto, J.B. Kogut and A. Kocic,
 Phys. Rev. {\bf D43} R1763 (1991).\hfill}
\vskip .1 truecm
\item {2.} {A. Kocic, J.B. Kogut, M.P. Lombardo and K.C. Wang, Spectroscopy,
 Scaling and Critical Indices in Strongly Coupled Quenched QED,
 { ILL-(TH)-92-12; CERN-TH.6542/92} (1992).\hfill}
\vskip .1 truecm
\item {3.} {V. Azcoiti, G. Di Carlo and A.F. Grillo, "A New Approach to
Non Compact Lattice QED with Light Fermions ",
 { DFTUZ 91.34} (1992).\hfill}
\vskip .1 truecm
\item {4.} {S.J. Hands, A. Kocic, J.B. Kogut, R.L. Renken,
D.K. Sinclair and K.C. Wang, "Spectroscopy, Equation of State and monopole
percolation in lattice QED with two flavours",
 {CERN-TH.6609/92} (1992).\hfill}
\vskip .1 truecm
\item {5.} {A. Kocic, J.B. Kogut and K.C. Wang,
" Monopole Percolation and the Universality Class of the Chiral
Transition in Four flavour Noncompact Lattice QED."
 ILL-TH-92-17 (1992).\hfill}
\vskip .1 truecm
\item {6.} {V. Azcoiti, G. Di Carlo and A.F. Grillo, "Renormalization Group
 and Triviality in Non Compact Lattice QED with Light Fermions",
 { DFTUZ 91.33} (1991), to appear in Mod. Phys. Lett. A \hfill}
\vskip .1 truecm
\item {7.} {M. G\"ockeler, R. Horsley, P. Rakow, G. Schierolz and R. Sommer,
Nucl. Phys. {\bf B371} 713 (1992)}
\vskip .1 truecm
\item {8.}{V. Azcoiti, G. Di Carlo and A.F. Grillo, Phys. Rev. Lett. {\bf 65}
2239 (1990). \hfill}
\vskip .1 truecm
\item {9.} { A. Kocic, J.B. Kogut and S.J. Hands,
"The Universality class of Monopole condensation in non compact, quenched QED"
 {ILL-(TH)-92-6} (1992).\hfill}
\vskip .1 truecm
\item {10.} {K.Kondo, Y. Kikukawa and H. Mino, Phys. Lett. {\bf B 220}
270 (1989)}
\item {11.}
{E. Dagotto, A. Kocic and J. B. Kogut, Phys. Lett. {\bf B 231} 235 (1989)}

\vfill
\eject

\line{}
\vskip 1 truecm
\centerline{\bf FIGURE CAPTIONS}
\vskip 1 truecm

\item {1)}{ First contribution to the effective fermionic action (Equation 8)
in a $8^4$ lattice, $m=0.0$ and $N=4$.
Errors are smaller than symbols.}

\vskip .1 truecm

\item {2)}{ Phase diagram in the $(\beta,N)$ plane at $m=0.0$. }

\vfill
\eject

\end